\documentclass[11pt,a4paper]{article}
\usepackage{graphicx} 
\usepackage{amsmath,amssymb,amsthm,bm,fullpage}
\usepackage{appendix}
\usepackage{booktabs}
\usepackage{caption}
\usepackage{rotating}
\usepackage{multirow}
\usepackage{authblk}
\usepackage{makecell}
\usepackage{threeparttable}
\usepackage[nottoc]{tocbibind}
\usepackage{color}
\usepackage{natbib}
\usepackage[pdftex]{hyperref}
\hypersetup{plainpages=True, pdfstartview=FitH, bookmarksopen=true,
colorlinks=true,linkcolor=blue,citecolor=blue}

\title{Efficient Variance Estimation for the Polytomous Discrimination Index}

\begin{document}
\author[1]{Qunqiang Feng}
\author[1]{Taozheng Jia}
\author[2]{Pan Liu}
\author[3]{Ben Van Calster}
\author[4,5]{Jialiang Li}

\affil[1]{Department of Statistics and Finance, School of Management, University of Science and Technology of China}
\affil[2]{Department of Statistics and Data Science, University of California, Los Angeles}
\affil[3]{Department of Development and Regeneration, KU Leuven, Belgium}
\affil[4]{Department of Statistics and Data Science, National University of Singapore}
\affil[5]{Duke-NUS Graduate Medical School, National University of Singapore}
\maketitle

\date{}

\begin{abstract}

Evaluating diagnostic accuracy for multi-category outcomes remains a significant challenge, 
primarily due to computational limitations in existing performance metrics. 
The Polytomous Discrimination Index (PDI) has emerged as an order-agnostic solution suitable for nominal classifications.
However, its broader adoption has been constrained by the lack of efficient implementation, 
especially for its variance estimation, which typically would require computationally intensive bootstrapping procedures. 
In this work, we address this limitation by proposing a novel asymptotic variance estimator for the PDI. 
Our method integrates classical $U$-statistic theory with recent advances in combinatorics, 
offering a scalable and theoretically grounded alternative. 
To assess the performance of the proposed approach, 
we conduct extensive simulation studies and observe remarkable gain in computing time. 
We further apply our method to a real-world brain image analysis where deep neural networks are used as a diagnostic tool. 
We can efficiently report the accuracy of the neural networks with different depth specifications.
 
\end{abstract}

\section{Introduction}\label{sec1}
Evaluating diagnostic and predictive accuracy is a cornerstone of medical research and clinical practice. Many real-world classification problems involve polytomous outcomes with more than two levels, such as distinct disease subtypes or multi-level severity scores~\cite{li2019evaluating,yang2020,li2025diagnostic}. In many cases, these outcomes are nominal, meaning they do not possess an inherent ordering. Representative examples include classifying influenza virus subtypes (e.g., H1N1, H3N2, H5N1) or distinguishing among brain tumor types (glioma, meningioma, pituitary tumor, or no tumor) from imaging data. Accurate classification is critical in these settings, motivating the widespread use of modern machine learning techniques, such as multinomial logistic regression, random forests, and convolutional neural networks. Accordingly, rigorous assessment of a model’s discriminatory power, its ability to correctly distinguish between outcome categories, is essential to determine whether the model is fit for clinical deployment or requires further refinement.

While the area under the receiver operating characteristic curve (AUC) serves as a standard discrimination measure for 
binary classification \cite{hanley1982,bradley1997,pepe2003statistical,krzanowski2009roc,zhou2011statistical,zou2011statistical,sande2021statistical,yang2024transformed}, 
its direct extension to nominal or ordinal outcomes with $M\ge 3$ categories faces significant limitations \cite{Mossman1999,ANY2009}. 
This gap has spurred the development of more structured frameworks for multi-class discrimination assessment. In particular the concepts of AUC was further extended to the volume under the ROC surface (VUS) 
for three-category outcomes \cite{Mossman1999,ANY2009} and the hypervolume under the ROC manifold (HUM) for arbitrary class counts \cite{NY2004,LF2008,WZW2011,lcw2014,feng2021discrete,maiti2023distribution,feng2023network,liu2024hypothesis,li2026hum}. 

The polytomous discrimination index (PDI), originally introduced by Van Calster et al. \cite{calstervvvs2012extending}, extended the concordance and C-statistics and provided another useful perspective for multi-category accuracy studies. In fact, the PDI quantifies the probability that a randomly selected subject from distinct classes is correctly classified \cite{li2018nonpara}. 
This measure is invariant to class labeling and accommodates non-transitive relationships \cite{calstervvvs2012extending,Ding2021}. Moreover,
it can be directly calculated by classification probability (risk) vector, making it well-suited for classification model validation of machine learning. PDI also reduces to the standard AUC for binary cases and now becomes a valuable tool for multi-class diagnostic evaluation \cite{Calstervlbts2012,ledger2023} and prediction accuracy \cite{Ding2021,Jiang2022,parmar2025}.

Despite its theoretical appeal, PDI adoption has been hindered by computational challenges. 
Existing implementations computing PDI include MATLAB-dependent R code
by Van Calster et al \cite{calstervvvs2012extending},
Gao and Li's earlier version of \texttt{pdi} function in the \texttt{mcca} R package \cite{gaoli2019},
and \%PDI SAS macro and R function by Dover et al \cite{dover2021computing}.
Although Dover et al. \cite{dover2021computing} allow linearithmic-time speed calculation for the point estimation of PDI in most large-scale datasets, 
their bootstrap-based variance calculation remains computationally intensive.
Consequently, reliable and efficient statistical inference is still challenging for complicated data sets such as the image data to be analyzed in this paper.

This paper bridges this gap by developing a computationally efficient method for the asymptotic variance estimation of PDI.  
Building on the classical $U$-statistic theory \cite{Lehmann1999} and leveraging the recent algorithmic advances \cite{dover2021computing,feng2023network}, 
we extend the prior works on nonparametric PDI estimation and inference \cite{li2018nonpara}. 
Our approach enables the confidence interval construction and hypothesis testing for PDI. The computational methods are now available in the updated version of \texttt{pdi} function in the  R package \texttt{mcca}.

The remainder of this paper is organized as follows. 
Section 2 reviews the definition of PDI and its standard estimation, 
followed by our proposed method for computing the asymptotic variance. 
We also present a toy example to illustrate the computational procedure and discuss implementation considerations. 
Section 3 demonstrates the performance of our method through extensive simulation studies.
In Section 4 we apply our methods to a real-world brain image dataset modeled by the convolutional neural network. Section 5 concludes with discussions and future directions.

\section{Methods}
\subsection{PDI and its estimation}

Let us begin by revisiting the definition of PDI \cite{calstervvvs2012extending}.
Consider a general diagnostic setting where the outcome is an $M$-category class variable with $M\ge2$ fixed.
The population version of PDI is defined as the probability that a randomly selected subject from a set of 
$M$ subjects (one from each category) is correctly classified as his true category \cite{li2018nonpara}. 
Formally, let $A_a$ denote the event that a randomly selected subject is from category $a$, 
and $B_a$ the event that this subject is classified into category $a$.
The probability definition of PDI is given by
\[
\theta=\sum_{a=1}^{M} P(B_a|A_a)P(A_a)=\sum_{a=1}^{M} \theta_a P(A_a),
\]
where $\theta_a=P(B_a|A_a)$ denotes the category-specific probability of correct classification for class $a$. 
In particular, if each category is selected with an equal probability (i.e., $P(A_a)\equiv 1/M$ for all $a$), 
then we have \cite{calstervvvs2012extending}
\begin{equation}\label{popPDI}
\theta= \frac{1}{M} \sum_{a=1}^{M}\theta_a.
\end{equation}
A naive random classifier assigns subjects to categories with an identical probability, yielding $\theta_a=1/M$ for all $a$.
Consequently, the null value of PDI is $1/M$ for such a random guess classifier.  

For any integer $1\le a\le M$,
define an $M$-variate function $C_a(\bm{v}_1,\dots,\bm{v}_M)$ for $M$-dimensional vector arguments $\bm{v}_b=(v_{1,b},\dots,v_{M,b})$, $b = 1, \dots, M$, as
\[
C_a(\bm{v}_1,\dots,\bm{v}_M)=\begin{cases}
	\frac1{m+1}, & \text{if $v_{a,a} \geq v_{a,b}$ for all $b\ne a$ and exactly $m$ equalities hold};\\
	0, & \text{otherwise}.
\end{cases}
\]
Here, $m=0,1,\dots,M-1,$ is the number of ties. This function is essentially comparing the $a$th element across the $M$ vectors. Such a comparison is frequently conducted for multi-class analysis where the input vectors $\bm{v}_b$ are usually the predicted probability vectors. Equivalently, 
\begin{equation}
	C_a(\bm{v}_1,\dots,\bm{v}_M)=\frac{\prod_{b \neq a} \bm{I}(v_{a,a} \geq v_{a,b})}{1+\sum_{b \neq a} \bm{I}(v_{a,a} = v_{a,b})},
\end{equation}
where $\bm{I}(\cdot)$ is the indicator function. The following construction will be based on this mathematical function.

The classification for multiple categories is most often based on the  predicted probabilities (or risk assessments) of the study subjects. Such risks
  are probability scores which are usually computed from individuals’ diagnostic test values or other biomarkers such as gene expressions, images,
  physical and biological measurements. Clinicians will make a decision for the patient's disease status based on these probabilities. 
  Let $\bm{p}_b=(p_{1,b},\dots,p_{M,b})$ denote the vector of predicted probabilities for a subject randomly selected from the $b$th category, $b=1,\dots,M$,
  where $p_{a,b}$ is the probability of assigning this subject to category $a$, satisfying $p_{a,b}\ge 0$ and $\sum_{a=1}^Mp_{a,b}=1$. Specifically, a subject from category $a$ is correctly classified only when $p_{a,a} \geq p_{a,b}$ for all $b\ne a$. 
  If exactly $m$ equalities hold, a random guess must be made, and the probability of correct classification becomes $1/(m+1)$.
  Therefore, for each $a=1,\dots,M$, we can express the correct classification probability $\theta_a$ as 
  \begin{equation*}
  	\theta_a
  	=\sum_{m=0}^{M-1}\frac1{m+1}P\Big(C_a(\bm{p}_1,\dots,\bm{p}_M)=\frac{1}{m+1}\Big),
  \end{equation*}
  which is equal to $\mathbb{E}[C_a(\bm{p}_1,\dots,\bm{p}_M)]$. 
  Consequently, by \eqref{popPDI} the overall PDI is given by
  \begin{equation}\label{popPDIpp}
  	\theta= \frac{1}{M} \sum_{a=1}^{M}\sum_{m=0}^{M-1}\frac1{m+1}P\Big(C_a(\bm{p}_1,\dots,\bm{p}_M)=\frac{1}{m+1}\Big).
  \end{equation}
  In particular, if no ties exist (i.e., $P(p_{a,a}=p_{a,b})=0$ for all $a\neq b$), 
  then \eqref{popPDIpp} reduces to
  \[
  \theta= \frac{1}{M} \sum_{a=1}^{M}P\Big(C_a(\bm{p}_1,\dots,\bm{p}_M)=1\Big),
  \]
  which aligns with the actionable definition of PDI given in \cite{li2018nonpara}.
  
We now consider the sample estimates of PDI. Suppose that we observe $n_a$ subjects from category $a$ for $a=1,\dots,M$, 
with a total sample size $n=\sum_{a=1}^Mn_a$.
For the $s$-th subject in category $b$ ($s=1,\dots,n_b$),
let $\bm{p}_{s,b}=(p_{1,s,b},p_{2,s,b},\dots,p_{M,s,b})$ denote the vector of predicted probabilities (or risk assessments),
where $p_{a,s,b}$ is the probability assigned to category $a$,
satisfying $p_{a,s,b}\ge0$ and $\sum_{a=1}^Mp_{a,s,b}=1$. 
We assume such a probability vector is available for all the sample subjects, 
typically derived from an upstream model fitting or training process using statistical or machine learning algorithms, 
such as logistic regression, decision trees, support vector machines, or neural networks.

The empirical nonparametric estimator of PDI  \cite{calstervvvs2012extending,dover2021computing} is constructed as follows. Note that the vectors $\{\bm{p}_{s,b},s=1,\dots,n_b\}$ constitutes a random i.i.d. sample from the distribution of $\bm{p}_b$, for each $b = 1, \dots, M$.
A nonparametric unbiased estimator of PDI for $M$ categories is given by \cite{calstervvvs2012extending,li2018nonpara,dover2021computing}
\begin{equation} \label{estPDI}
	\hat{\theta}=\frac{1}{M} \sum_{a=1}^{M} \hat{\theta}_a,
\end{equation}
where $\hat{\theta}_a$ is the corresponding estimator of $\theta_a$ given by
\begin{equation} \label{pe1}
	\hat{\theta}_a=\frac{1}{\prod_{b=1}^M n_b} \sum_{s_1=1}^{n_1} \cdots \sum_{s_M=1}^{n_M}
	C_a(\bm{p}_{s_1,1},\dots,\bm{p}_{s_M,M}), \quad a=1,\dots,M.
\end{equation}

The computational complexity of direct implementation of \eqref{estPDI} is of order $O(\prod_{a=1}^Mn_a)$ .
This approach becomes prohibitive for large $M$ and/or sample sizes. To address this problem,
Dover et al. \cite{dover2021computing} proposed an equivalent expression that enables linearithmic-time computation in sample size. 
For each category $a$, define the set of distinct predicted probabilities
\[\mathcal{S}_a= \{p_{a,s_a,a}: s_a= 1,\dots, n_a\}.\] 
For any $p\in[0,1]$ and category $b$, 
let $n_{a,p,b}$ and $N_{a,p,b}$ denote the number of subjects whose predicted 
probability for category $a$ equal to $p$, and smaller than $p$, respectively. That is, for any $b=1,\dots,M$,
\begin{align*}
  n_{a,p,b}=\left|\{p_{a,s_b,b}:p_{a,s_b,b}=p, s_b=1,\dots,n_b\}\right|, \\  
  N_{a,p,b}=\left|\{p_{a,s_b,b}:p_{a,s_b,b}<p, s_b=1,\dots,n_b\}\right|, 
\end{align*}
where $|\cdot|$ denotes set cardinality. 
Using these notations, \eqref{pe1} can be rewritten as \cite{dover2021computing}
\begin{equation}\label{pe2}
    \hat{\theta}_a = \frac{1}{\prod_{b=1}^M n_b} \sum_{p \in \mathcal{S}_a} \sum_{\mathcal{C}_a}\frac{1}{\sum_{b=1}^M c_b} \prod_{b=1}^M \big(c_b n_{a,p,b} +(1-c_b) N_{a,p,b}\big),
\end{equation}
where the inner sum is over the set of $M$-dimensional vectors
\[\mathcal{C}_a=\{(c_1,\dots,c_M): c_a=1, c_b\in \{0,1\} \text{ for } b\neq a\}.\]
Therefore, the overall PDI estimator \eqref{estPDI} can be rewritten as
\begin{equation} \label{estPDIDover}
\hat{\theta}=\frac{1}{M\prod_{b=1}^M n_b}\sum_{a=1}^M
\sum_{p \in \mathcal{S}_a} \sum_{\mathcal{C}_a}\frac{1}{\sum_{b=1}^M c_b} \prod_{b=1}^M \big(c_b n_{a,p,b} +(1-c_b) N_{a,p,b}\big).
\end{equation}
A naive computation of the quantities $n_{a,p,b}$ and $N_{a,p,b}$
would require repeatedly scanning the data, leading to quadratic complexity. 
However, as shown in \cite{dover2021computing}  and implemented in \cite{Ding2021}, 
these counts can be obtained efficiently by first sorting the predicted probabilities (cost $O(n\log n)$)
and then calculating cumulative counts recursively in linear time for each category. 
After this preprocessing, the summation in \eqref{estPDIDover} involves $O(2^Mn)$  terms, 
each taking constant time to evaluate. 
Hence, the overall computational complexity of computing $\hat{\theta}$ via \eqref{estPDIDover} is $O(n\log n+2^Mn)$.
In practice, $M$ is often small, so the factor $2^M$ acts as a constant and the method scales 
near-linearithmically with sample size.

Note that if no ties occur at each $p\in \mathcal{S}_a$,
the counts $n_{a,p,a}=1$ and $n_{a,p,b}=0$ for all $b\neq a$. 
We can then simplify  \eqref{pe2} to
\begin{equation} \label{pe2notie}
\hat{\theta}_a = \frac{1}{\prod_{b=1}^M n_b} \sum_{p \in \mathcal{S}_a} \prod_{b \neq a} N_{a,p,b}, \quad a=1,\dots,M,
\end{equation}
and \eqref{estPDIDover} to
\begin{equation*} 
\hat{\theta}=\frac{1}{M\prod_{b=1}^M n_b}\sum_{a=1}^M
\sum_{p \in \mathcal{S}_a} \prod_{b \neq a} N_{a,p,b}.   
\end{equation*}

\subsection{Computing the variance of estimated PDI}

In order to conduct statistical inference, estimating the variance of the sample estimator $\hat{\theta}$ is essential in practice. 
From expression \eqref{estPDI}, we can easily show that the exact variance is
\[
 \mathrm{Var}(\hat{\theta})=\frac{1}{M^2}\sum_{a=1}^{M} \sum_{b=1}^{M}\mathrm{Cov}
 \big(\hat{\theta}_a,\hat{\theta}_b\big). 
\]
Thus, calculation of the covariance between $\hat{\theta}_a$ and $\hat{\theta}_b$ for each pair $(a,b)$ is required.
However, direct computation of these covariances using \eqref{pe2} or \eqref{pe2notie} remains challenging due to 
the large number of product terms involving counting variables.
To address this issue, we combine the $U$-statistics theory with the combinatorial framework \cite{dover2021computing}  
to develop the asymptotic variance formula.

We first note that the estimator in \eqref{estPDI} is an $M$-sample $U$-statistic with the kernel \cite{Lehmann1999} 
\[
\phi(\bm{p}_1,\dots,\bm{p}_M)=\frac1M\sum_{a=1}^MC_a(\bm{p}_1,\dots,\bm{p}_M), 
\]
which has exactly one single argument per sample.
By the classical $U$-statistics theory \cite{Lehmann1999}, the estimator is consistent to the true value of PDI and asymptotically normal, 
as all the sample sizes $n_1,\dots,n_M$ tend to infinity.
To formalize the asymptotic normality, define for $d=1,\dots,M$, 
\begin{align}\label{sigmad2}
\sigma_d^2 :&= \mathrm{Cov} \left (\phi(\bm{p}_1,\dots,\bm{p}_M),
\phi(\bm{p}_1',\cdots,\bm{p}_{d-1}',\bm{p}_d,\bm{p}_{d+1}',\cdots,\bm{p}_M')\right ) \notag \\
&= \mathrm{Cov} \left (\frac{1}{M} \sum_{a=1}^M C_a(\bm{p}_1,\dots,\bm{p}_M),
\frac{1}{M} \sum_{b=1}^M C_b\big(\bm{p}_1',\cdots,\bm{p}_{d-1}',\bm{p}_d,\bm{p}_{d+1}',\cdots,\bm{p}_M'\big)\right ), 
\end{align}
where $\bm{p}_b'$  denotes an independent copy of $\bm{p}_{b}$
for any $b=1,\dots,d-1,d+1,\dots,M$.
Assume $n_1,\dots,n_M \to\infty$ with limiting proportions
\[
\rho_i=\lim_{n\to\infty}\frac{n_i}{n}\in (0,1).
\]
By Theorem 6.1.4 of \cite{Lehmann1999},  if $\sigma_1^2,\dots,\sigma_M^2>0$, 
\[
\sqrt{n} \thinspace (\hat{\theta}-\theta) \xrightarrow{d} N(0,\sigma^2),
\]
where $\xrightarrow{d}$ denotes convergence in distribution,
and the asymptotic variance is
\begin{equation}
    \sigma^2 = \sum_{d=1}^M \frac{\sigma_d^2}{\rho_d}.
    \label{sigma}
\end{equation}
A 95\% confidence interval for PDI can then be constructed as
$$
\hat{\theta} \pm 1.96 \thinspace \hat{\sigma}/\sqrt{n},
$$
if we can obtain a consistently estimated asymptotic variance $\hat{\sigma}^2$ from the sample.

Let us try to work out more details for the variance term in order to derive a sensible estimator. Since $\mathbb{E}[C_a(\bm{p}_1,\dots,\bm{p}_M)]=\theta_a$, 
by \eqref{popPDI} and \eqref{sigmad2} we have 
\begin{align}\label{sigmad22}
  \sigma_d^2 
    & = \mathbb{E} \left(
   \frac{1}{M} \sum_{a=1}^M C_a(\bm{p}_1,\dots,\bm{p}_M) \cdot
\frac{1}{M} \sum_{b=1}^M C_b\big(\bm{p}_1',\cdots,\bm{p}_{d-1}',\bm{p}_d,\bm{p}_{d+1}',\cdots,\bm{p}_M'\big) \right) - \theta^2 \notag\\
    & = \frac{1}{M^{2}}\sum_{a=1}^M\sum_{b=1}^Mq_d^{(ab)} - \theta^2,  
\end{align}
where 
\begin{equation}\label{qdab}
q_d^{(ab)}=\mathbb{E}\big(C_a(\bm{p}_1,\dots,\bm{p}_M)\cdot C_b\big(\bm{p}_1',\cdots,\bm{p}_{d-1}',\bm{p}_d,\bm{p}_{d+1}',\cdots,\bm{p}_M'\big)\big), \quad a,b=1,\dots,M.
\end{equation}
Combining \eqref{sigma} and \eqref{sigmad22} yields
\begin{equation}\label{sigmad22new}
   \sigma^2 = \frac{1}{M^{2}} \sum_{d=1}^M\sum_{a=1}^M\sum_{b=1}^M\frac{q_d^{(ab)}}{\rho_d}
              -\theta^2 \sum_{d=1}^M\frac{1}{\rho_d},
\end{equation}
which agrees to the result in \cite{li2018nonpara} for the special case $M=3$ in the absence of ties.
Noting that in \eqref{qdab}, all arguments except the $d$th are independent replicates in the functions $C_a$ and $C_b$, 
by the law of total expectation we can proceed with
\begin{align}\label{qdab1}
q_d^{(ab)}&=\mathbb{E}\big[\mathbb{E}\big(C_a(\bm{p}_1,\dots,\bm{p}_M)\cdot C_b\big(\bm{p}_1',\cdots,\bm{p}_{d-1}',\bm{p}_d,\bm{p}_{d+1}',\cdots,\bm{p}_M'\big)\big)\big|\bm{p}_{d}\big]\notag\\
&=\mathbb{E}\big[\mathbb{E}\big(C_a(\bm{p}_1,\dots,\bm{p}_M)\big|\bm{p}_{d} \big)\cdot
\mathbb{E}\big(C_b\big(\bm{p}_1',\cdots,\bm{p}_{d-1}',\bm{p}_d,\bm{p}_{d+1}',\cdots,\bm{p}_M'\big)\big|
\bm{p}_{d}\big)\big],
\end{align}
where the last equality holds due to independence. 

We now derive the empirical estimates of $q_d^{(ab)}$ for all $a,b$ and $d$ using the observed data. 
A key insight here is that the term $\mathbb{E}(C_a(\bm{p}_1,\dots,\bm{p}_M)|\bm{p}_{d})$ corresponds to $\theta_{a|d}$, the probability that a random subject from category $a$ is correctly classified, when the probability vector of a subject randomly selected from category $d$ is fixed at $\bm{p}_{d}$. 
Following this, we can rewrite \eqref{qdab1} as 
\begin{align*}
q_d^{(ab)}=\mathbb{E}\big(\theta_{a|d} \cdot \theta_{b|d}\big), \quad a,b,d=1,\dots,M.
\end{align*}
Leveraging the empirical sample and analogous to \eqref{pe2},
the estimate of $q_d^{(ab)}$ can be calculated by
\begin{align}\label{hatqdab} 
\hat{q}_d^{(ab)}=\frac{1}{n_d}\sum_{s_d=1}^{n_d}\hat{\theta}_{a|(s_d,d)}\cdot\hat{\theta}_{b|(s_d,d)},
\end{align} 
where 
\begin{align}
\hat{\theta}_{a|(s_d,d)}&= \frac{n_d}{\prod_{b=1}^Mn_b} \sum_{p \in \mathcal{S}_a} \sum_{\mathcal{C}_a}
\frac{c_d\bm{I}(p=p_{a,s_d,d})+(1-c_d)\bm{I}(p>p_{a,s_d,d})}{\sum_{b=1}^Mc_b}\notag\\
&\quad\times\prod_{b\ne d} \big(c_b n_{a,p,b} +(1-c_b) N_{a,p,b}\big)\label{hatPDIasdd}
\end{align}
can be regarded as the estimator of $\theta_a$ 
for the subsample retaining only one subject in category $d$ (whose index is $s_d$) while keeping the other categories unaltered. 
We provide a concise explanation of \eqref{hatPDIasdd} as follows.
Setting $n_d=1$ in \eqref{pe2} yields the factor $n_d/\prod_{b=1}^Mn_b$ in \eqref{hatPDIasdd}. 
Comparing the product terms in \eqref{pe2} and \eqref{hatPDIasdd},
 we can show that $n_{a,p,d}=\bm{I}(p=p_{a,s_d,d})$
and $N_{a,p,d}=\bm{I}(p>p_{a,s_d,d})$ for this scenario. For any $d=1,\dots,M$ and $s_d=1,\dots,n_d$, 
analogous to \eqref{estPDI} we define the subsample PDI estimate
\begin{align}\label{PDIsdd}
\hat{\theta}_{(s_d,d)}=\frac1M\sum_{a=1}^M\hat{\theta}_{a| (s_d,d)}.
\end{align} 
In particular, if there exist no ties in the predicted probabilities, 
$\hat{\theta}_{a|(s_d,d)}$ given in \eqref{hatPDIasdd} can be rewritten as 
\begin{align}
\hat{\theta}_{a|(s_d,d)}=\frac{n_d}{\prod_{b=1}^Mn_b} \cdot \begin{cases}
 \sum_{p \in \mathcal{S}_a} \bm{I}(p>p_{a,s_d,d})\prod_{b\ne a,d} N_{a,p,b},& \mbox{if~} d\ne a;\\
   \bm{I}(p=p_{a,s_d,d})\prod_{b\ne d}N_{a,p,b},& \mbox{if~} d = a,
 \end{cases}
\end{align}
and thus \eqref{PDIsdd} can be reduced to 
\begin{align}\label{PDIsddnotie}
\hat{\theta}_{(s_d,d)}=
 \frac{n_d}{M\prod_{b=1}^Mn_b}\left(\bm{I}(p=p_{d,s_d,d})\prod_{b\ne d} N_{d,p,b}+
 \sum_{a\ne d} \sum_{p \in \mathcal{S}_a} \bm{I}(p>p_{a,s_d,d})\prod_{b\ne a,d} N_{a,p,b}
 \right).
\end{align}

 Finally, based on \eqref{sigmad22new},
 the estimator $\hat{\sigma}^2$  is then obtained by substituting empirical counterparts $\hat{q}_d^{(ab)}$
 and $\hat{\theta}$
 into \eqref{sigmad22new},  along with the sample proportions $\hat{\rho}_d=n_d/n$. 
 Thus,  combining \eqref{hatqdab} and \eqref{PDIsdd} gives that
\begin{align}
\hat{\sigma}^2
&=\frac{1}{M^{2}} \sum_{d=1}^M\sum_{a=1}^M\sum_{b=1}^M\frac{n}{n_d^2}\sum_{s_d=1}^{n_d}\hat{\theta}_{a|(s_d,d)}\cdot\hat{\theta}_{b|(s_d,d)}
              -n\hat{\theta}^2\sum_{d=1}^M \frac{1}{n_d}\notag\\
&=n\sum_{d=1}^M\frac1{n_d^2}\sum_{s_d=1}^{n_d}\left(\frac{1}{M^2}
\sum_{a=1}^M\sum_{b=1}^M\hat{\theta}_{a|(s_d,d)}
\cdot\hat{\theta}_{b|(s_d,d)}\right)-n\hat{\theta}^2\sum_{d=1}^M \frac{1}{n_d}
\notag\\
&=n\sum_{d=1}^M\frac1{n_d^2}\sum_{s_d=1}^{n_d}\hat{\theta}_{(s_d,d)}^2
-n\hat{\theta}^2\sum_{d=1}^M \frac{1}{n_d}\label{hatsigma2}\\
&=n\sum_{d=1}^M\frac{1}{n_d}\left(\frac{1}{n_d}\sum_{s_d=1}^{n_d}\hat{\theta}_{(s_d,d)}^2-\hat{\theta}^2\right). \label{hatsigma3}
\end{align} 
Notice that for any $d=1,\dots,M$,
\begin{align} \label{ndpdisdd}
\frac{1}{n_d}\sum_{s_d=1}^{n_d}\hat{\theta}_{(s_d,d)}=\hat{\theta},
\end{align}
which implies that each term in the parentheses on the right-hand side of 
\eqref{hatsigma3} is not less than 0 for $d=1,\dots,M$.
Likewise, we also have that for any $a,d=1,\dots,M$,
\begin{align} \label{ndpdiasdd}
\frac{1}{n_d}\sum_{s_d=1}^{n_d}\hat{\theta}_{a|(s_d,d)}=\hat{\theta}_a.
\end{align}
Thus,  \eqref{ndpdisdd} and \eqref{ndpdiasdd} provide an alternative way to calculate $\hat{\theta}$ and $\hat{\theta}_{a}$ for any $a=1,\dots,M$, respectively.

Equation \eqref{hatsigma2} or \eqref{hatsigma3}, combined with  \eqref{ndpdisdd},
indicates that estimating the asymptotic variance $\sigma^2$ only requires calculating
the estimates  $\hat{\theta}_{(s_d,d)}$ for all $d=1,\dots,M$ and $s_d=1,\dots,n_d$. Consequently, the computational complexity of calculating $\hat{\sigma}^2$ through 
\eqref{hatsigma2} is of order $O(n^2)$.

Alternatively, combining \eqref{hatPDIasdd}, \eqref{PDIsdd} and \eqref{hatsigma2} gives that
\begin{align}\label{hatsigma22}
\hat{\sigma}^2&=\frac{n}{(M\prod_{b=1}^Mn_b)^2}\sum_{d=1}^M\sum_{s_d=1}^{n_d}
\left(\sum_{a=1}^M\sum_{p \in \mathcal{S}_a} \sum_{\mathcal{C}_a}
\frac{c_d\bm{I}(p=p_{a,s_d,d})+(1-c_d)\bm{I}(p>p_{a,s_d,d})}{\sum_{b=1}^Mc_b}\right.\notag\\
&\qquad\left.\times\prod_{b\ne d} \big(c_b n_{a,p,b} +(1-c_b) N_{a,p,b}\big)\right)^2
 -n\hat{\theta}^2\sum_{d=1}^M\frac{1}{n_d}.
\end{align}
In particular, for the case with no ties, by \eqref{PDIsddnotie} simplifications can be conducted on \eqref{hatsigma22} into 
\begin{align} \label{hatsigma2notie}
\hat{\sigma}^2&=\frac{n}{(M\prod_{b=1}^Mn_b)^2}\sum_{d=1}^M\sum_{s_d=1}^{n_d}\left(\bm{I}( p=p_{d,s_d,d}) \prod_{b\ne d} N_{d,p,b}+\sum_{a\ne d}\sum_{p \in \mathcal{S}_a} 
\bm{I}(p>p_{a,s_d,d}) \prod_{b\ne a,d} N_{a,p,b} 
\right)^2\notag\\
&\quad -n\hat{\theta}^2\sum_{d=1}^M\frac{1}{n_d}.
\end{align} 
Since the double sums in \eqref{hatsigma22} and \eqref{hatsigma2notie} can be regarded as
the Frobenius norm of a $|\mathcal{S}_1\bigcup\cdots\bigcup\mathcal{S}_M|\times n$ matrix,
the computational complexity of our algorithm using these two equations is still of order $O(n^2)$. 

\subsection{Toy example}
 To illustrate the computation of the estimator of asymptotic variance $\sigma^2$ using our computational method, 
we consider a toy example with $M=3$ categories and $n=8$ subjects ($n_1=1,n_2=4,n_3=3$).
The predicted probabilities are shown in Table \ref{tab:toy}.  
This toy dataset is adapted from Dover et al. \cite{dover2021computing}. 
While the original source provides predicted probabilities only for class 1, 
we supplement such probabilities for classes 2 and 3 to construct a complete example.
\begin{table}[htbp]
\centering
\caption{\label{toy}Toy data ($n_1=1,n_2=4,n_3=3$)}
\begin{tabular}{@{}ccc@{}}
\toprule
Subject & Category & Predicted Probabilities\\
\midrule
1 & 1 & $(0.4,  0.3, 0.3)$ \\
2 & 2 & $(0.1,  0.5, 0.4)$ \\
3 & 2 & $(0.3,  0.3, 0.4)$ \\
4 & 2 & $(0.4,  0.4, 0.2)$ \\
5 & 2 & $(0.5,  0.4, 0.1)$ \\
6 & 3 & $(0.2,  0.3, 0.5)$ \\
7 & 3 & $(0.4,  0.2, 0.4)$ \\
8 & 3 & $(0.4,  0.1, 0.5)$ \\
\bottomrule
\end{tabular}\label{tab:toy}
\end{table}

To better present the idea of our method, 
we apply \eqref{hatsigma2} rather than \eqref{hatsigma22} in the following. 

 First, the sets of distinct probabilities are
\[
\mathcal{S}_1=\{0.4\}, \quad \mathcal{S}_2=\{0.3,0.4,0.5\}, \quad \mathcal{S}_3=\{0.4,0.5\}.
\]
For all categories $a,d=1,2,3$ and predicted probabilities $p\in \mathcal{S}_a$, 
the counts $n_{a,p,d}$ and $N_{a,p,d}$ are computed from Table \ref{tab:toy} and summarized in Table \ref{tab:counts}.
Since $c_a$ is identically 1 for all vectors in $\mathcal{C}_a$,  
the coefficient $1-c_a$ of $N_{a,p,a}$ in \eqref{hatPDIasdd} is always zero. 
Consequently, calculating $N_{a,p,a}$ for each $1\le a\le M$ is unnecessary in our algorithm (see also Dover et al. \cite{dover2021computing}).
Therefore, values of $N_{a,p,a}$ are omitted from Table \ref{tab:counts}. 

\begin{table}[htbp]
\centering
\caption{Count variables $n_{a,p,d}$ and $N_{a,p,d}$.}
\label{tab:counts}
\begin{threeparttable}
\begin{tabular}{cccccccc}
  \toprule
  \multirow{2}{*}{}&\multirow{2}{*}{$p$}&\multicolumn{3}{c}{$n_{a,p,d}$}&\multicolumn{3}{c}{$N_{a,p,d}$}\\
  \cmidrule(lr){3-5} \cmidrule(lr){6-8}
  ~&~&$d=1$&$d=2$&$d=3$&$d=1$&$d=2$&$d=3$\\
  \midrule
  $a=1$ & 0.4&1&1&2&-&2&1\\
  \specialrule{0em}{3pt}{3pt}
  $a=2$&0.3&1&1&1&0&-&2\\
  &0.4&0&2&0&1&-&3\\
  &0.5&0&1&0&1&-&3\\
  \specialrule{0em}{3pt}{3pt}
  $a=3$&0.4&0&2&1&1&2&-\\
  &0.5&0&0&2&1&4&-\\
  \bottomrule
\end{tabular}
\begin{tablenotes}
\item -: value omitted.   
\end{tablenotes}
\end{threeparttable}
\end{table}

Next, we compute the estimates $\hat{\theta}_{(s_d,d)}$ for all $s_d$ and $d$. 
Notice that by \eqref{PDIsdd}, $\hat{\theta}_{(s_d,d)}$ equals the average of estimates $\hat{\theta}_{a|(s_d,d)}$ over $a$.
For illustration, the computations for $\hat{\theta}_{1|(1,1)}$ and $\hat{\theta}_{2|(1,1)}$  via \eqref{hatPDIasdd}
are detailed below (others are analogous). 

Noting that $\mathcal{S}_1$ contains a single element $p=0.4$ and $c_1=1$ and accounting for $c_2,c_3\in\{0,1\}$,
\eqref{hatPDIasdd} and  Table \ref{tab:counts} yield that
\begin{align*}
    \hat{\theta}_{1|(1,1)}&=\frac{1}{n_2 n_3}\left(\frac{1}{1+1+1}n_{1,0.4,2}n_{1,0.4,3}+\frac{1}{1+1+0} n_{1,0.4,2}N_{1,0.4,3}\right. \notag\\
    &\quad \left.+\frac{1}{1+0+1} N_{1,0.4,2} n_{1,0.4,3}
    + \frac{1}{1+0+0}N_{1,0.4,2} N_{1,0.4,3}\right)\\
    &=\frac{1}{4\cdot 3}\cdot\left(\frac13\cdot1\cdot2+\frac12\cdot1\cdot1+\frac12\cdot2\cdot2+1\cdot2\cdot1\right)\\
    &=\frac{31}{72}.
\end{align*}
Since there is only one subject in category 1,  
it follows by \eqref{ndpdiasdd} that $\hat{\theta}_{1|(1,1)}=\hat{\theta}_{1}$, the estimate of the category-specific term ${\theta}_1$,  consistent with Dover et al. \cite{dover2021computing}.

For $\hat{\theta}_{2|(1,1)}$ with $\mathcal{S}_2=\{0.3,0.4,0.5\}$ and $c_2=1$,
note that sum terms on the right-hand side of \eqref{hatPDIasdd} vanish when
$c_d\bm{I}(p=p_{a,s_d,d})+(1-c_d)\bm{I}(p>p_{a,s_d,d})=0$ (here $a=2,d=1$).
Given $p_{2,1,1}=0.3$ (Table \ref{tab:toy}), 
the term $c_1\bm{I}(p=0.3)+(1-c_1)\bm{I}(p>0.3)$ is not equal to 0 (and thus equal to 1) 
only in two cases: (1) $c_1=1$ for $p=0.3$, and (2) $c_1=0$ for $p=0.4$ or 0.5.
Accounting for $c_3$ being 1 or 0, by \eqref{hatPDIasdd} and  Table \ref{tab:counts} we thus have 
\begin{align*}
    \hat{\theta}_{2|(1,1)}&=
    \frac{1}{n_2 n_3}
    \left(\frac{1}{1+1+1} n_{2,0.3,2} n_{2,0.3,3}+\frac{1}{1+1+0} n_{2,0.3,2} N_{2,0.3,3}\right.\\
    &\quad +\frac{1}{0+1+1} n_{2,0.4,2} n_{2,0.4,3}+\frac{1}{0+1+0} n_{2,0.4,2} N_{2,0.4,3}\\
    &\quad \left.+\frac{1}{0+1+1} n_{2,0.5,2}n_{2,0.5,3}+\frac{1}{0+1+0} n_{2,0.5,2} N_{2,0.5,3} \right) \\
    &=\frac{1}{4\cdot 3}\cdot\left(\frac13\cdot1\cdot1+\frac12\cdot1\cdot2+\frac12\cdot2\cdot0+1\cdot2\cdot3+\frac12\cdot1\cdot0+1\cdot1\cdot3\right)\notag\\
    &=\frac{31}{36}. 
\end{align*}
All $\hat{\theta}_{a|(s_d,d)}$ and $\hat{\theta}_{(s_d,d)}$  values are summarized in Table \ref{tab:pdiasdd}. 
\begin{table}[htbp]
\centering
\caption{Subsample PDI estimates $\hat{\theta}_{a|(s_d,d)}$ and $\hat{\theta}_{(s_d,d)}$.}
\label{tab:pdiasdd}
\begin{tabular}{cccccc}
  \toprule
  \multicolumn{2}{c}{\multirow{2}{*}{$(s_d,d)$}}&\multicolumn{3}{c}{$\hat{\theta}_{a|(s_d,d)}$}&\multirow{2}{*}{$\hat{\theta}_{(s_d,d)}$}\\
  \cmidrule(lr){3-5}
  \multicolumn{2}{c}{}&$a=1$&$a=2$&$a=3$&~\\
  \midrule
  \multicolumn{2}{c}{$(1,1)$}&31/72&31/36&11/12&53/72\\
  \specialrule{0em}{3pt}{3pt}
  \multicolumn{2}{c}{$(1,2)$}&2/3&1&5/6&5/6\\
  \multicolumn{2}{c}{$(2,2)$}&2/3&4/9&5/6&35/54\\
  \multicolumn{2}{c}{$(3,2)$}&7/18&1&1&43/54\\
  \multicolumn{2}{c}{$(4,2)$}&0&1&1&2/3\\
  \specialrule{0em}{3pt}{3pt}
  \multicolumn{2}{c}{$(1,3)$}&5/8&5/6&1&59/72\\
  \multicolumn{2}{c}{$(2,3)$}&1/3&7/8&3/4&47/72\\
  \multicolumn{2}{c}{$(3,3)$}&1/3&7/8&1&53/72\\
  \bottomrule
\end{tabular}
\end{table}

Finally, we apply \eqref{hatsigma2} and \eqref{ndpdisdd} to compute the estimate of the variance $\hat{\sigma}^2$.
It follows by \eqref{ndpdisdd} that the PDI estimate of our toy dataset is the average of $\hat{\theta}_{(s_d,d)}$ over $s_d$, 
independent of $d$.
Then Table \ref{tab:pdiasdd} gives that 
\begin{align}\label{toyPDI}
\hat{\theta}=1\cdot\frac{53}{72}=\frac14\cdot\Big(\frac56+\frac{35}{54}+\frac{43}{54}+\frac{2}{3}\Big)=
\frac{1}{3}\cdot\Big(\frac{59}{72}+\frac{47}{72}+\frac{53}{72}\Big)=\frac{53}{72}.
\end{align}
(Alternatively, one can compute $\hat{\theta}$ by using \eqref{ndpdiasdd} and \eqref{estPDI}. 
In fact, \eqref{ndpdiasdd} and Table \ref{tab:pdiasdd} yield that
\begin{align*}
\hat{\theta}_{1}&=1\cdot \frac{31}{72}=\frac{1}{4}\cdot\Big(\frac23+\frac23+\frac{7}{18}+0\Big)
                         =\frac13\cdot\Big(\frac58+\frac13+\frac13\Big)=\frac{31}{72},\\
\hat{\theta}_{2}&=1\cdot \frac{31}{36}=\frac{1}{4}\cdot\Big(1+\frac49+1+1\Big)
                         =\frac13\cdot\Big(\frac56+\frac78+\frac78\Big)=\frac{31}{36},\\ 
\hat{\theta}_{3}&=1\cdot \frac{11}{12}=\frac{1}{4}\cdot\Big(\frac56+\frac56+1+1\Big)
                         =\frac13\cdot\Big(1+\frac34+1\Big)=\frac{11}{12}.
\end{align*}
Then \eqref{estPDI} further gives 
\[
\hat{\theta}=\frac{1}{3}\cdot\Big(\frac{31}{72}+\frac{31}{36}+\frac{11}{12}\Big)=\frac{53}{72}=0.7361,
\]
leading to the same result as given in \eqref{toyPDI}.)
Therefore,  by \eqref{hatsigma2} and Table \ref{tab:pdiasdd}, we arrive at 
\begin{align*}
    \hat{\sigma}^2&=8\cdot \left[ \frac{53^2}{72^2}+\frac{1}{4^2}\cdot \left(\frac{5^2}{6^2}+\frac{35^2}{54^2}+\frac{43^2}{54^2}+\frac{2^2}{3^2}\right)+\frac{1}{3^2}\cdot \left(\frac{59^2}{72^2}+\frac{47^2}{72^2}+\frac{53^2}{72^2}\right) \right]\\
    &\quad-8\cdot \frac{53^2}{72^2}\cdot \left(1+\frac{1}{4}+\frac{1}{3}\right)\\
    &=0.0252,
\end{align*}
and the asymptotic standard deviation of our PDI estimator for the toy dataset is
\[
\frac{\hat{\sigma}}{\sqrt{n}}=\sqrt{\frac{0.0252}{8}}=0.0561.
\]
The 95\% confidence interval can then be constructed as $0.7361\pm 1.96\times0.0561=(0.6261,0.8460)$.

\section{Simulation Study}
In this section, we conduct a simulation study to evaluate the performance of our proposed method 
for computing the asymptotic standard error estimator of PDI. 
Our method is benchmarked against the bootstrap approach by Dover et al. \cite{dover2021computing} 
to assess estimation accuracy and computational efficiency.
Following the ADEMP structure \cite{morris2019}, we have summarized relevant information of this simulation study in Table \ref{tab: ademp}.

\begin{table}[htbp]
	\centering
	\caption{ADEMP structure for the simulation design.}
	\begin{tabular}{l|l}
		\hline
		Aims  & \makecell[l]{$\bullet$ Compute PDI estimates.\\
			$\bullet$ Compute SE of PDI estimation using our proposed method.\\
			$\bullet$ Compare the accuracy with the bootstrap method. \\
			$\bullet$ Compare the computational efficiency with the bootstrap method.} \\ \hline
		Data generation & 
		\makecell[l]{$\bullet$ Generate biomarker values from 8 different specified distributions
			\\ ~~ and fit a multinomial logistic regression model to obtain \\
			~~ the corresponding predictive probability vectors. \\
			$\bullet$ Generate predictive probability vectors from Dirichlet
			distributions  \\ ~~ and multinomial distributions.
		}\\ \hline
		Estimands & \makecell[l]{$\bullet$  PDI with and without tied observations. \\
			$\bullet$ SE of estimated PDI with and without tied observations.}\\ \hline
		Methods &  
		\makecell[l]{$\bullet$ Estimation method in \cite{dover2021computing} for PDI.\\
			$\bullet$ Our estimators for SE of estimated PDI (Subsection 2.2). \\
			$\bullet$ Bootstrap method for SE of estimated PDI. \cite{dover2021computing}
		}\\ \hline
		Performance measures & 
		\makecell[l]{$\bullet$ Mean absolute errors (MAE).\\
			$\bullet$ Average computing time (seconds).
		}\\
		\hline
	\end{tabular}\label{tab: ademp}
	\begin{tablenotes}
		\footnotesize
		\item Abbreviations: 
		SE, standard error;
		MAE, average of absolute error.
	\end{tablenotes}
\end{table}

Data are generated using two distinct strategies:
\begin{itemize}
\item[(1)] Biomarker-based generation: Biomarkers are simulated from specified univariate distributions, 
followed by fitting a multinomial logistic regression model to obtain the corresponding predictive probability vectors.
\item[(2)] Direct probability generation: 
Predictive probability vectors for distinct categories are sampled directly 
from specified multivariate distributions.
\end{itemize}

Both approaches include scenarios with and without ties, 
allowing evaluation of \eqref{hatsigma22} (with ties) and \eqref{hatsigma2notie} (no ties). 
For each experiment, we perform 500 simulations with category sizes
$n+\mathrm{Bin}(n/2,0.5)$ for $n=80,200,400,1000$, 
where $\mathrm{Bin}(n/2,0.5)$ denotes a binomial variate with parameters $n/2$ and $0.5$.
Computational cost is quantified by the average real elapsed time per calculation across simulations.
All simulations are performed on a PC with Intel(R) Core(TM) i9-12900K 3.20 GHz processor and 64.0 GB RAM.

\begin{figure}[htbp]
	\centering
	\includegraphics[width=0.9\textwidth]{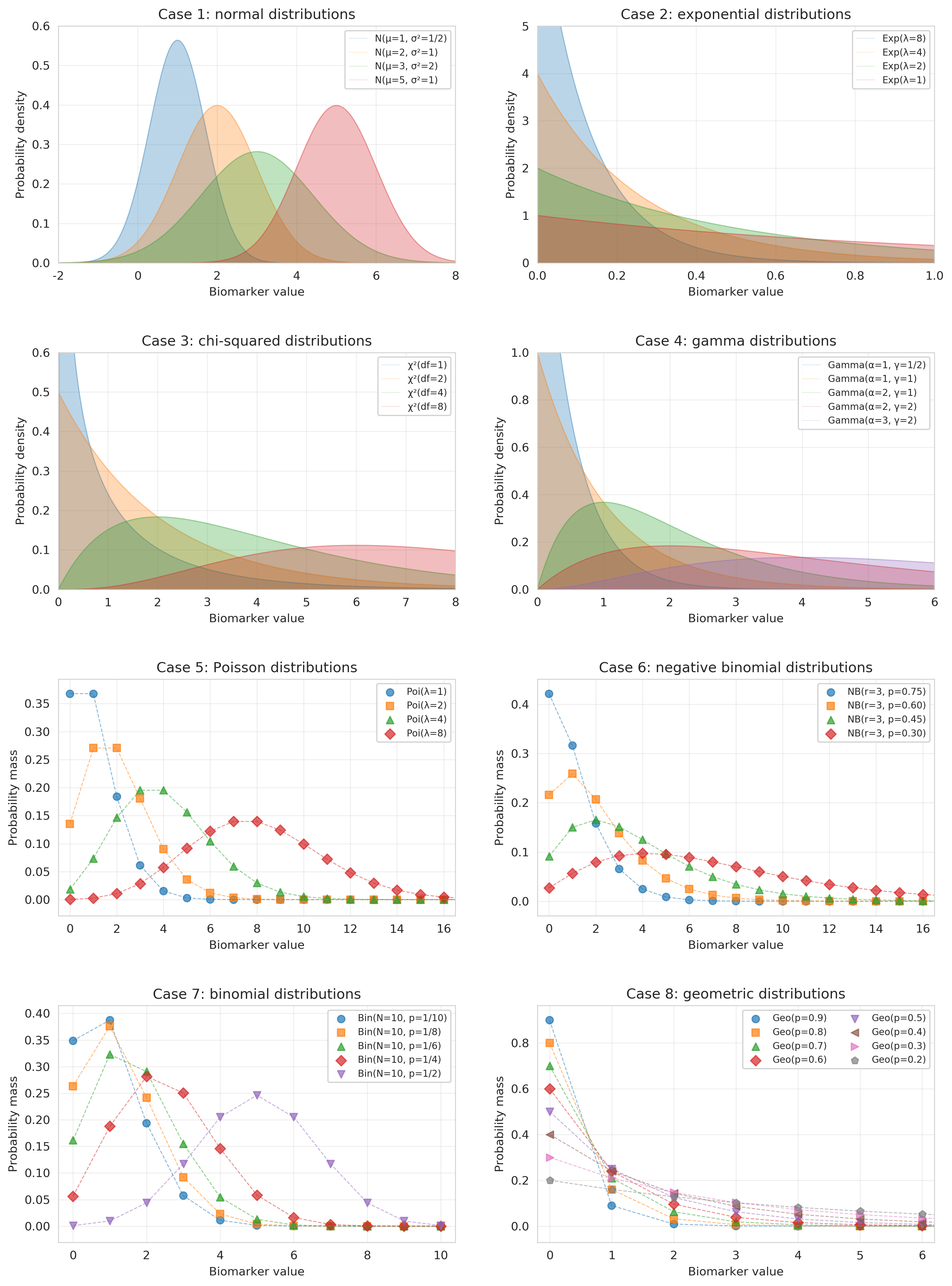}
	\caption{Probability distributions of the generated biomarkers (Cases 1--8).}
	\label{bio_distribution}
\end{figure}

We first look into eight scenarios featuring biomarkers from well-known distributions (Figure \ref{bio_distribution}). 
Continuous distributions (Cases 1–4) generate minimal ties, 
while discrete distributions (Cases 5–8) produce frequent ties. 
Parameterizations ensure distinct separation across categories.

\begin{itemize}
\item[] \textbf{Case 1} ($M=4$): 
Biomarkers are sampled from four normal distributions with means 1, 2, 3, 5, 
and variances 0.5, 1, 2, 1, respectively.

\item[] \textbf{Case 2} ($M=4$): 
Biomarkers are sampled from the exponential distributions
$$
\text{Exp}(\lambda_d):\quad f_d(x)=\lambda_d e^{-\lambda_d x} \quad (x \geq 0)
$$
with the parameter $\lambda_d=8,4,2,1$ for $d=1,2,3,4$, respectively.
Note that distribution $\text{Exp}(\lambda)$ has a mean of $1/\lambda$.

\item[] \textbf{Case 3} ($M=4$):
Biomarkers are sampled from the chi-squared distributions
$$
\chi^2(k_d):\quad f_d(x)=\dfrac{1}{2^{k_d/2} \Gamma(k_d/2)} x^{k_d/2-1} e^{-x/2} \quad (x \geq 0)
$$
with degree of freedom $k_d=1,2,4,8$ for $d=1,2,3,4$, respectively. 
Note that distribution $\chi^2(k)$ has a mean of $k$.

\item[] \textbf{Case 4} ($M=5$): 
Biomarkers are sampled from the gamma distributions
$$
\Gamma(\alpha_d ,\gamma_d):\quad f_d(x)=\dfrac{1}{\Gamma(\alpha_d) \gamma_d^{\alpha_d} x^{\alpha_d-1} e^{-x/\alpha_d}} \quad (x \geq 0)
$$
with the shape parameter $\alpha_d = 1, 1, 2, 2, 3$ and the scale parameter $\gamma_d = 0.5, 1, 1, 2, 2$ for $d = 1, 2, 3, 4, 5$, respectively. We
note that distribution $\Gamma(\alpha ,\gamma)$ has a mean of $\alpha \gamma$.

\item[] \textbf{Case 5} ($M=4$):  Biomarkers are sampled from Poisson distributions
$$\text{Poi}(\lambda_d):\quad f_d(x)=\dfrac{\lambda_d^x}{x!} e^{-x} \quad (x=0,1,2,\cdots)$$
with the parameter $\lambda_d=1,2,4,8$ for $d=1,2,3,4$, respectively. 
Note that distribution $\text{Poi}(\lambda)$ has a mean of $\lambda$.

\item[] \textbf{Case 6} ($M=4$): 
Biomarkers are sampled from negative binomial distributions 
$$
\text{NB}(r,p_d):\quad f_d(x)=\binom{x+r-1}{r-1} p_d^r (1-p_d)^{x} \quad (x=0,1,2,\cdots)
$$
with the number of successes $r = 3$ and success probability $p_d = 0.75,0.60,0.45,0.30$ for $d = 1, 2, 3, 4$, respectively. 
Note that distribution $\text{NB}(r,p)$ has a mean of $r(1-p)/p$.

\item[] \textbf{Case 7} ($M=5$): 
Biomarkers are sampled from the binomial distributions
$$\text{Bin}(N,p_d):\quad f_d(x)=\binom{N}{x} p_d^x (1-p_d)^{(N-x)} \quad (x = 0,1,\cdots,N)$$
with constant number of trials $N = 10$ and the incremental success probabilities 
$p_d = 1/10,1/8,1/6, 1/4,1/2$ 
for $d = 1,\dots, 5$, respectively.

\item[] \textbf{Case 8} ($M=8$): 
Biomarkers are sampled from the geometric distributions
$$
\text{Geo}(p_d):\quad f_d(x)=p_d (1-p_d)^{x} \quad (x=0,1,2,\cdots)
$$
with the success probability $p_d=0.9,0.8,\dots,0.2$ for $d=1,\dots,8$, respectively.
Note that distribution $\text{Geo}(p)$ has a mean of $(1-p)/p$.
\end{itemize}

The simulation results for Cases 1–8 are summarized in Table \ref{simbio}.
\begin{table}[htbp]
\small
\centering
\setlength{\tabcolsep}{4pt}
\caption{Simulation results for Cases 1--8.}
\vspace{5pt}
\label{simbio}
\begin{threeparttable}
\begin{tabular}{lcccccccccccr}
  \toprule
  \multirow{2}{*}{Case}
  &\multirow{2}{*}{$M$}&\multirow{2}{*}{Class sample sizes}&\multirow{2}{*}{$\hat{\theta}$}&\multirow{2}{*}{SD}&\multicolumn{4}{c}{Our method}&\multicolumn{4}{c}{Bootstrap method}\\
    \cmidrule(r){6-9} \cmidrule(r){10-13}
    &&&&&SE&MAE&CV&Time(s)&SE&MAE&CV&Time(s)\\
  \midrule
  1 & 4&80+Bin(40,.5)&0.6369&0.0229&0.0214&0.0015&0.93&0.0132&0.0214&0.0019&0.93&4.1744\\
  &&200+Bin(100,.5)&0.6350&0.0135&0.0134&0.0002&0.94&0.0471&0.0134&0.0008&0.94&10.8924\\
  &&400+Bin(200,.5)&0.6343&0.0095&0.0096&0.0002&0.93&0.1269&0.0097&0.0006&0.93&21.7497\\
  &&1000+Bin(500,.5)&0.6340&0.0063&0.0061&0.0001&0.95&0.5502&0.0062&0.0004&0.95&60.1264\\
  &&100,150,300,500&0.6167&0.0173&0.0155&0.0018&0.91&0.0497&0.0154&0.0019&0.91&11.9270\\
  \specialrule{0em}{3pt}{3pt}
  
   2 &4&80+Bin(40,.5)&0.4592&0.0223&0.0212&0.0012&0.94&0.0138&0.0213&0.0017&0.94&4.3111\\
  &&200+Bin(100,.5)&0.4593&0.0135&0.0133&0.0003&0.94&0.0463&0.0132&0.0009&0.94&11.2417\\
  &&400+Bin(200,.5)&0.4581&0.0091&0.0095&0.0004&0.95&0.1297&0.0095&0.0007&0.95&21.0357\\
  &&1000+Bin(500,.5)&0.4585&0.0059&0.0060&0.0001&0.95&0.5558&0.0060&0.0003&0.95&61.7543\\
  &&100,150,300,500&0.4376&0.0122&0.0118&0.0009&0.93&0.0492&0.0117&0.0011&0.93&10.8497\\
  \specialrule{0em}{3pt}{3pt}
  
3 & 4&80+Bin(40,.5)&0.5621&0.0233&0.0221&0.0012&0.92&0.0136&0.0221&0.0016&0.92&4.2423\\
  &&200+Bin(100,.5)&0.5633&0.0143&0.0139&0.0004&0.95&0.0468&0.0140&0.0008&0.95&11.1630\\
  &&400+Bin(200,.5)&0.5634&0.0100&0.0099&0.0001&0.94&0.1383&0.0099&0.0006&0.94&22.7814\\
  &&1000+Bin(500,.5)&0.5648&0.0062&0.0063&0.0001&0.96&0.5571&0.0062&0.0004&0.96&60.8397\\
  &&100,150,300,500&0.5583&0.0155&0.0144&0.0011&0.93&0.0515&0.0144&0.0011&0.93&11.8497\\

  \specialrule{0em}{3pt}{3pt}
  
  4 & 5&80+Bin(40,.5)&0.4679&0.0200&0.0189&0.0011&0.93&0.0187&0.0189&0.0016&0.93&12.9229\\
  &&200+Bin(100,.5)&0.4686&0.0121&0.0118&0.0003&0.95&0.0655&0.0117&0.0007&0.95&34.2253\\
  &&400+Bin(200,.5)&0.4673&0.0086&0.0085&0.0001&0.96&0.1961&0.0086&0.0005&0.96&66.8788\\
  &&1000+Bin(500,.5)&0.4674&0.0056&0.0053&0.0002&0.94&0.8364&0.0054&0.0004&0.94&188.9232\\
  \specialrule{0em}{3pt}{3pt}
  
  5 &4&80+Bin(40,.5)&0.6349&0.0213&0.0218&0.0006&0.94&0.0115&0.0221&0.0016&0.95&0.5634\\
  &&200+Bin(100,.5)&0.6325&0.0137&0.0136&0.0002&0.94&0.0218&0.0136&0.0008&0.94&0.8094\\
  &&400+Bin(200,.5)&0.6319&0.0107&0.0097&0.0011&0.93&0.0388&0.0097&0.0011&0.93&1.1747\\
  &&1000+Bin(500,.5)&0.6312&0.0065&0.0061&0.0004&0.93&0.1002&0.0061&0.0005&0.93&2.3336\\
  &&100,150,300,500&0.6040&0.0165&0.0141&0.0024&0.91&0.0217&0.0143&0.0022&0.91&0.8456\\
  \specialrule{0em}{3pt}{3pt}

  6 &4&80+Bin(40,.5)&0.4913&0.0224&0.0213&0.0012&0.95&0.0196&0.0212&0.0017&0.95&1.0969\\
  &&200+Bin(100,.5)&0.4908&0.0133&0.0134&0.0003&0.95&0.0410&0.0135&0.0009&0.95&1.5191\\
  &&400+Bin(200,.5)&0.4911&0.0095&0.0096&0.0002&0.95&0.0828&0.0095&0.0006&0.95&1.9997\\
  &&1000+Bin(500,.5)&0.4902&0.0060&0.0060&0.0001&0.95&0.2275&0.0060&0.0004&0.95&3.2942\\
  &&100,150,300,500&0.4644&0.0125&0.0112&0.0015&0.93&0.0434&0.0111&0.0016&0.93&1.6803\\
  \specialrule{0em}{3pt}{3pt}

  7 &5&80+Bin(40,.5)&0.4414&0.0177&0.0157&0.0020&0.92&0.0259&0.0158&0.0020&0.92&1.6892\\
  &&200+Bin(100,.5)&0.4416&0.0101&0.0097&0.0005&0.94&0.0518&0.0097&0.0007&0.94&2.1093\\
  &&400+Bin(200,.5)&0.4416&0.0070&0.0069&0.0001&0.95&0.1112&0.0069&0.0004&0.95&2.7602\\
  &&1000+Bin(500,.5)&0.4418&0.0044&0.0044&0.0001&0.95&0.2447&0.0044&0.0002&0.95&4.7238\\
  \specialrule{0em}{3pt}{3pt}

  8&8&80+Bin(40,.5)&0.2260&0.0082&0.0081&0.0003&0.94&0.4131&0.0081&0.0005&0.94&17.6338\\
  &&200+Bin(100,.5)&0.2251&0.0055&0.0052&0.0003&0.93&0.8824&0.0052&0.0004&0.93&21.1288\\
  &&400+Bin(200,.5)&0.2247&0.0039&0.0037&0.0002&0.94&1.9213&0.0037&0.0003&0.94&25.1386\\
  &&1000+Bin(500,.5)&0.2249&0.0023&0.0023&0.0001&0.97&5.4218&0.0023&0.0001&0.97&32.9010\\
  \bottomrule
\end{tabular}
\begin{tablenotes}
\footnotesize
\item Abbreviations: $M$, the number of categories;
$\hat{\theta}$, 
the average of estimated PDI values across 500 experiments; SD, the standard deviation of $\hat{\theta}$ across 500 experiments;
SE, the average of standard error estimates over 100 simulations;
MAE, the average of absolute error of SE over 100 simulations;
CV, coverage probability of the 95\% confidence interval for $\theta$ constructed by $[\hat{\theta}-1.96\thinspace \text{SE},\hat{\theta}+1.96\thinspace \text{SE}]$ in 100 experiments;
Time(s), average computation cost measured by the real elapsed time (in seconds) for computing the standard error over 100 simulations.
\end{tablenotes}
\end{threeparttable}
\end{table}

Next, we consider the predicted probability vectors that are sampled from 
the Dirichlet distribution (Cases 9--11; continuous, minimal ties)
and multinomial distribution (Cases 12--14; discrete, frequent ties).
The probability density function of the Dirichlet distribution with parameters $\bm{\alpha}=(\alpha_1,\dots,\alpha_M)$ is given by
\[
\text{Dir}(\bm{\alpha}):\quad f(x_1,\cdots,x_M)=\dfrac{\Gamma(\alpha_1+\cdots+\alpha_M)}{\Gamma(\alpha_1)\cdots \Gamma(\alpha_M)} \prod_{i=1}^M x_i^{\alpha_i-1}, 
\]
where $x_i >0$ and $\sum_{i=1}^M x_i=1$, 
while the probability mass function of the multinomial distribution with 
parameters $N$ and $\bm{p}=(p_1,\dots,p_M)$ ($p_i>0$ and $\sum_{i=1}^M p_i=1$) is 
\[
\text{Mul}(N,\bm{p}):\quad 
f(x_1,\cdots,x_{M})=\dfrac{N!}{x_1! \cdots x_M!} p_1^{x_1} p_2^{x_2} \cdots p_M^{x_M},
\]
where $x_1,\cdots,x_{M}$ are non-negative integers and $x_1+\cdots+x_{M} = N$. 
Vectors sampled from $\text{Mul}(N,\bm{p})$ are divided by $N$ to obtain probability vectors.
Here we set $N=10$. 
We also note that the mean vectors of distributions 
$\text{Dir}(\bm{\alpha})$ 
and $\text{Mul}(N,\bm{p})$
are
$(\alpha_1/\sum_{i=1}^M\alpha_i,\dots,\alpha_M/\sum_{i=1}^M\alpha_i)$
and $(Np_1,\dots,Np_M)$, respectively.

\begin{itemize}
\item[] \textbf{Case 9} ($M=4$): $\bm{\alpha}=(1,2,3,4)$ for all classes.

\item[]\textbf{Case 10} ($M=4$): 
$\bm{\alpha}_1=(4,3,2,1)$,
$\bm{\alpha}_2=(1,4,3,2)$, 
$\bm{\alpha}_3=(2,1,4,3)$, and
$\bm{\alpha}_4=(3,2,1,4)$.

\item[] \textbf{Case 11} ($M=5$): 
$\bm{\alpha}_1=(1,2,3,4,5)$,
$\bm{\alpha}_2=(1,2,3,2,1)$, 
$\bm{\alpha}_3=(5,4,3,2,1)$,
$\bm{\alpha}_4=(2,1,3,5,4)$, and 
$\bm{\alpha}_5=(1,1,1,1,1)$.

\item[] \textbf{Case 12} ($M=4$): $\bm{p}=(0.25,0.20,0.15,0.40)$ for all classes.

\item[] \textbf{Case 13} ($M=4$): 
$\bm{p}_1=(0.30,0.25,0.20,0.25)$,
$\bm{p}_2=(0.25,0.40,0.15,0.20)$, 
$\bm{p}_3=(0.20,0.35,0.30,0.15)$, and
$\bm{p}_4=(0.15,0.30,0.25,0.30)$.

\item[] 
\textbf{Case 14} ($M=5$): 
 For class $i$, the vector $\bm{p}_i$ has $i$-th entry $1/3$, and other entries $1/6$. 
\end{itemize}

The simulation results for Cases 9--14 are listed in Table \ref{simpro}.
\begin{table}[htbp]
\small
\centering
\setlength{\tabcolsep}{4pt}  
\caption{Simulation results for Cases 9--14.}
\vspace{5pt}
\label{simpro}
\begin{threeparttable}
\begin{tabular}{lcccccccccccr}
\toprule
  \multirow{2}{*}{Case}
  &\multirow{2}{*}{$M$}&\multirow{2}{*}{Sample sizes}&\multirow{2}{*}{$\hat{\theta}$}&\multirow{2}{*}{SD}&\multicolumn{4}{c}{Our method}&\multicolumn{4}{c}{Bootstrap method}\\
    \cmidrule(r){6-9} \cmidrule(r){10-13}
    &&&&&SE&MAE&CV&Time(s)&SE&MAE&CV&Time(s)\\
  \midrule
  9 & 4&80+Bin(40,.5)&0.2511&0.0187&0.0189&0.0007&0.95&0.0115&0.0189&0.0014&0.95&4.2279\\
  &&200+Bin(100,.5)&0.2494&0.0117&0.0117&0.0003&0.95&0.0462&0.0116&0.0007&0.95&11.1659\\
  &&400+Bin(200,.5)&0.2504&0.0081&0.0084&0.0003&0.96&0.1435&0.0083&0.0006&0.96&22.3887\\
  &&1000+Bin(500,.5)&0.2500&0.0054&0.0053&0.0001&0.94&0.6002&0.0054&0.0003&0.94&63.0956\\
  &&100,150,300,500&0.2510&0.0137&0.0134&0.0005&0.95&0.0458&0.0134&0.0010&0.95&11.2914\\
  \specialrule{0em}{3pt}{3pt}
  
  10 & 4&80+Bin(40,.5)&0.6217&0.0249&0.0243&0.0006&0.95&0.0129&0.0244&0.0014&0.95&4.1684\\
  &&200+Bin(100,.5)&0.6229&0.0151&0.0150&0.0001&0.95&0.0454&0.0149&0.0008&0.95&10.9523\\
  &&400+Bin(200,.5)&0.6226&0.0109&0.0107&0.0002&0.95&0.1426&0.0106&0.0007&0.95&22.0996\\
  &&1000+Bin(500,.5)&0.6227&0.0067&0.0068&0.0001&0.95&0.5872&0.0067&0.0004&0.95&61.3851\\
  &&100,150,300,500&0.6234&0.0183&0.0177&0.0007&0.96&0.0425&0.0176&0.0013&0.96&11.1192\\
  \specialrule{0em}{3pt}{3pt}
  
  11 & 5&80+Bin(40,.5)&0.2025&0.0146&0.0141&0.0006&0.94&0.0182&0.0141&0.0011&0.94&12.8091\\
  &&200+Bin(100,.5)&0.2019&0.0085&0.0087&0.0002&0.96&0.0707&0.0087&0.0005&0.96&33.8855\\
  &&400+Bin(200,.5)&0.2019&0.0061&0.0062&0.0001&0.95&0.2026&0.0062&0.0003&0.95&69.4709\\
  &&1000+Bin(500,.5)&0.2021&0.0039&0.0039&0.0001&0.95&0.8518&0.0040&0.0002&0.95&194.1767\\
  \specialrule{0em}{3pt}{3pt}
  
  12 & 4&80+Bin(40,.5)&0.2501&0.0173&0.0185&0.0012&0.97&0.0178&0.0184&0.0017&0.97&0.9063\\
  &&200+Bin(100,.5)&0.2500&0.0109&0.0114&0.0005&0.96&0.0375&0.0112&0.0007&0.96&1.2438\\
  &&400+Bin(200,.5)&0.2499&0.0085&0.0081&0.0004&0.93&0.0704&0.0081&0.0005&0.93&1.7523\\
  &&1000+Bin(500,.5)&0.2502&0.0054&0.0051&0.0003&0.94&0.1755&0.0051&0.0004&0.94&3.2471\\
  &&100,150,300,500&0.2504&0.0139&0.0133&0.0007&0.95&0.0327&0.0134&0.0009&0.95&1.0858\\
  \specialrule{0em}{3pt}{3pt}
  
  13 & 4&80+Bin(40,.5)&0.4484&0.0234&0.0232&0.0004&0.95&0.0188&0.0233&0.0013&0.95&0.8918\\
  &&200+Bin(100,.5)&0.4481&0.0144&0.0144&0.0001&0.95&0.0358&0.0144&0.0009&0.95&1.2257\\
  &&400+Bin(200,.5)&0.4482&0.0099&0.0103&0.0004&0.96&0.0678&0.0102&0.0006&0.96&1.7168\\
  &&1000+Bin(500,.5)&0.4478&0.0069&0.0065&0.0005&0.93&0.1802&0.0065&0.0006&0.93&3.1889\\
  &&100,150,300,500&0.4479&0.0173&0.0170&0.0003&0.95&0.0365&0.0170&0.0011&0.95&1.2420\\
  \specialrule{0em}{3pt}{3pt}

  14 & 5&80+Bin(40,.5)&0.5754&0.0204&0.0204&0.0002&0.95&0.0320&0.0204&0.0012&0.95&1.2231\\
  &&200+Bin(100,.5)&0.5749&0.0124&0.0127&0.0003&0.95&0.0567&0.0126&0.0007&0.95&1.8172\\
  &&400+Bin(200,.5)&0.5750&0.0088&0.0091&0.0003&0.97&0.1288&0.0091&0.0006&0.97&2.9243\\
  &&1000+Bin(500,.5)&05749&0.0059&0.0057&0.0001&0.95&0.3130&0.0057&0.0003&0.95&4.9299\\
  \bottomrule
\end{tabular}
\begin{tablenotes}
\footnotesize
\item Abbreviations: $M$, the number of categories;
$\hat{\theta}$, 
the average of estimated PDI values across 500 experiments; SD, the standard deviation of $\hat{\theta}$ across 500 experiments;
SE, the average of standard error estimates over 100 simulations;
MAE, the average of absolute error of SE over 100 simulations;
CV, coverage probability of the 95\% confidence interval for $\theta$ constructed by $[\hat{\theta}-1.96\thinspace \text{SE},\hat{\theta}+1.96\thinspace \text{SE}]$ in 100 experiments;
Time(s), average computation cost measured by the real elapsed time (in seconds) for computing the standard error over 100 simulations.
\end{tablenotes}
\end{threeparttable}
\end{table}

From Tables \ref{simbio} and \ref{simpro}, 
one can observe that our method achieves variance estimates indistinguishable 
from bootstrap (i.e., SE $\approx$ SE.bs) while reducing computation time significantly
(by 2--4 orders of magnitude),  particularly for large $M$ or sample size $n$.

\subsection{Software implementation}

To facilitate practical application, we have incorporated the proposed variance estimation method into the \texttt{mcca} R package, specifically in the function \texttt{pdi()}. This subsection demonstrates its usage using a simulated dataset generated under Case 1. Users need to first install the \texttt{mcca} package from CRAN and load it:

\begin{verbatim}
install.packages("mcca")
library(mcca)
\end{verbatim}

The \texttt{pdi()} function is designed to be highly flexible, accommodating two primary types of input data: raw continuous biomarker values or a pre-computed predicted probability matrix. We first illustrate the use of raw biomarker values. For the generated dataset containing $n$ subjects, the function requires two main inputs. The first is a numeric vector \texttt{y} of length $n$, where the $i$-th element represents the integer category label for the $i$-th subject. The second is a numeric vector \texttt{d\_biom}, also of length $n$, which stores the corresponding raw biomarker value for each subject. The first six entries of our simulated data are as follows:

\begin{verbatim}
head(y)
## Output:
## [1] 4 3 4 2 3 1

head(d_biom)
## Output:
## [1] 4.260857 3.583994 7.475202 3.074347 4.075989 1.423995
\end{verbatim}

To compute the PDI and its asymptotic variance, we call the \texttt{pdi()} function with the argument \texttt{Var = TRUE}:

\begin{verbatim}
result1 = pdi(y = y, d = d_biom, Var = TRUE, withTies = FALSE)
# or equivalently
# result1 = pdi(y = y, d = d_biom, Var = TRUE, method = "multinom", withTies = FALSE)

print(result1)

## Output:
## Call:
## pdi(y = y, d = d_biom, Var = TRUE, withTies = FALSE)
## 
## Overall Polytomous Discrimination Index:
##  0.5870177 
## 
## Standard Error:
##  0.01465413 
## 
## 95% Confidence Interval:
##  [0.5583, 0.6157]
## 
## Category-specific Polytomous Discrimination Index:
##  CATEGORIES    VALUES         SE  LOWER_CI  UPPER_CI
##           1 0.7447654 0.02896877 0.6879876 0.8015431
##           2 0.3409104 0.03217679 0.2778450 0.4039757
##           3 0.3750827 0.02576758 0.3245792 0.4255862
##           4 0.8873124 0.01253748 0.8627394 0.9118854
\end{verbatim}

In this function call, the default setting \texttt{method = "multinom"} is used, which automatically applies the multinomial logistic regression as the underlying classifier in computing PDI. However, users can easily specify other built-in classifiers, including classification trees (\texttt{method = "tree"}), support vector machines (\texttt{method = "svm"}), and linear discriminant analysis (\texttt{method = "lda"}). Because Case 1 features continuous biomarkers that produce minimal ties, we also set \texttt{withTies = FALSE} to trigger the optimized formula and maximize computational efficiency. However, if the data is believed to follow discrete distributions that frequently produce tied probabilities (as in Case 5), users should specify \texttt{withTies = TRUE} to accommodate the ties using our generalized variance formula. 

As shown above, the function outputs the overall PDI estimate alongside the asymptotic standard error and confidence intervals computed using our method. It also outputs category-specific components, their asymptotic standard errors, and confidence intervals in a structured table.

In addition to receiving raw biomarker values, the \texttt{pdi()} function also accepts a directly supplied predicted probability matrix. This feature is particularly useful if users have already obtained classification probabilities from an external classifier of their choice. To illustrate this, we use the predicted probability matrix \texttt{d\_prob} obtained by fitting the multinomial logistic regression to \texttt{y} and \texttt{d\_biom}. To meet the input requirements, \texttt{d\_prob} must be formatted as an $n \times M$ dataframe, where $M$ represents the total number of categories. The element in the $i$-th row and $j$-th column should correspond to the predicted probability that the $i$-th subject belongs to the $j$-th category. A preview of this dataframe is shown below:

\begin{verbatim}
head(d_prob)
## Output:
##           X1           X2         X3          X4
## 2.034079e-03 4.484200e-02 0.27781763 0.675306286
## 1.038124e-02 1.184454e-01 0.47847755 0.392695788
## 9.955703e-08 5.009536e-05 0.00236532 0.997584485
## 2.804184e-02 1.948474e-01 0.57041455 0.206696163
## 3.285477e-03 6.050467e-02 0.33353076 0.602679098
## 2.583619e-01 3.603001e-01 0.37179454 0.009543414
\end{verbatim}

When using \texttt{y} and \texttt{d\_prob} as the input, we need to explicitly set the argument \texttt{method = "prob"}. This informs the function that the \texttt{d} argument contains a pre-computed probability matrix rather than raw biomarker values. We call the function as follows:

\begin{verbatim}
result2 = pdi(y = y, d = d_prob, Var = TRUE, method = "prob", withTies = FALSE)
\end{verbatim}

This alternative call yields exactly the same output as \texttt{result1}.

\section{Brain Tumor MRI Data Analysis}

\begin{figure}[htbp]
    \centering
    \includegraphics[width=0.95\textwidth]{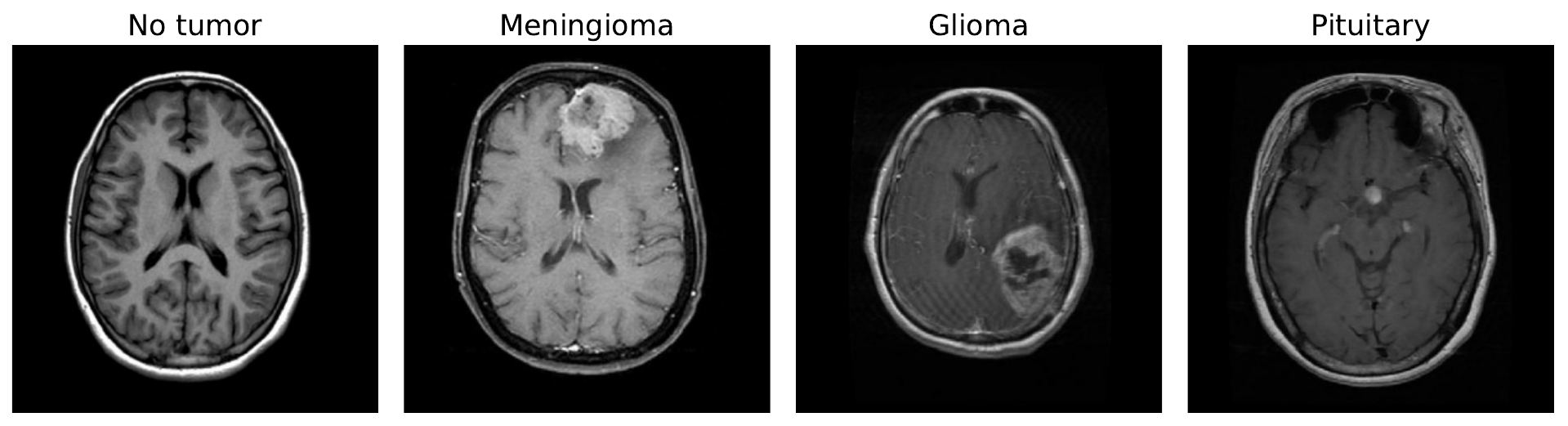}
    \caption{Axial view at the center of the brain MRI for four representative subjects.}
    \label{fig:type}
\end{figure}

We now illustrate our proposed methodology using a multi-source brain tumor MRI dataset, publicly available at the following website:
\begin{center}
 \url{https://www.kaggle.com/datasets/masoudnickparvar/brain-tumor-mri-dataset}.   
\end{center}
A brain tumor refers to an abnormal growth of cells within the skull, which can be either benign or malignant. As the tumor expands, it increases intracranial pressure, potentially leading to brain damage and life-threatening complications. The dataset used in this study comprises 7,023 magnetic resonance images (MRI) from four distinct tumor categories:
Glioma (cancerous brain tumors arising from glial cells), 
Meningioma (non-cancerous tumors originating from the meninges), 
No tumor (normal brain scans without detectable lesions), 
Pituitary (tumors affecting the pituitary gland, which can be cancerous or non-cancerous). 
Figures \ref{fig:type} displays typical individuals from each of the four categories in the dataset.  These medical images enable clinicians to visualize disease progression and support diagnostic decision-making, an area of increasing relevance in recent biomedical research. Traditionally, manual interpretation of brain scans requires extensive clinical training, is subject to observer variability, and is labor-intensive. Recent advances in artificial intelligence (AI) and deep learning \cite{goodfellow2016deep}, coupled with the availability of large annotated datasets, have enabled the development of automated systems for objective and scalable tumor classification to support clinicians. The goal of this study is to classify the four brain tumor types using a deep learning–based model and evaluate its diagnostic accuracy using the Polytomous Discrimination Index (PDI).

We conducted the image-based classification by adopting the convolutional neural network (CNN) architectures.
CNN is a specialized class of deep learning models designed for processing grid-like data such as images. A CNN architecture typically involves the following main parts: convolutional layers for feature extraction, pooling layers for spatial dimensionality reduction, and fully-connected layers for final classification.

\begin{figure}[ht]
    \centering
    \includegraphics[height=0.64\textheight,width=0.95\textwidth]{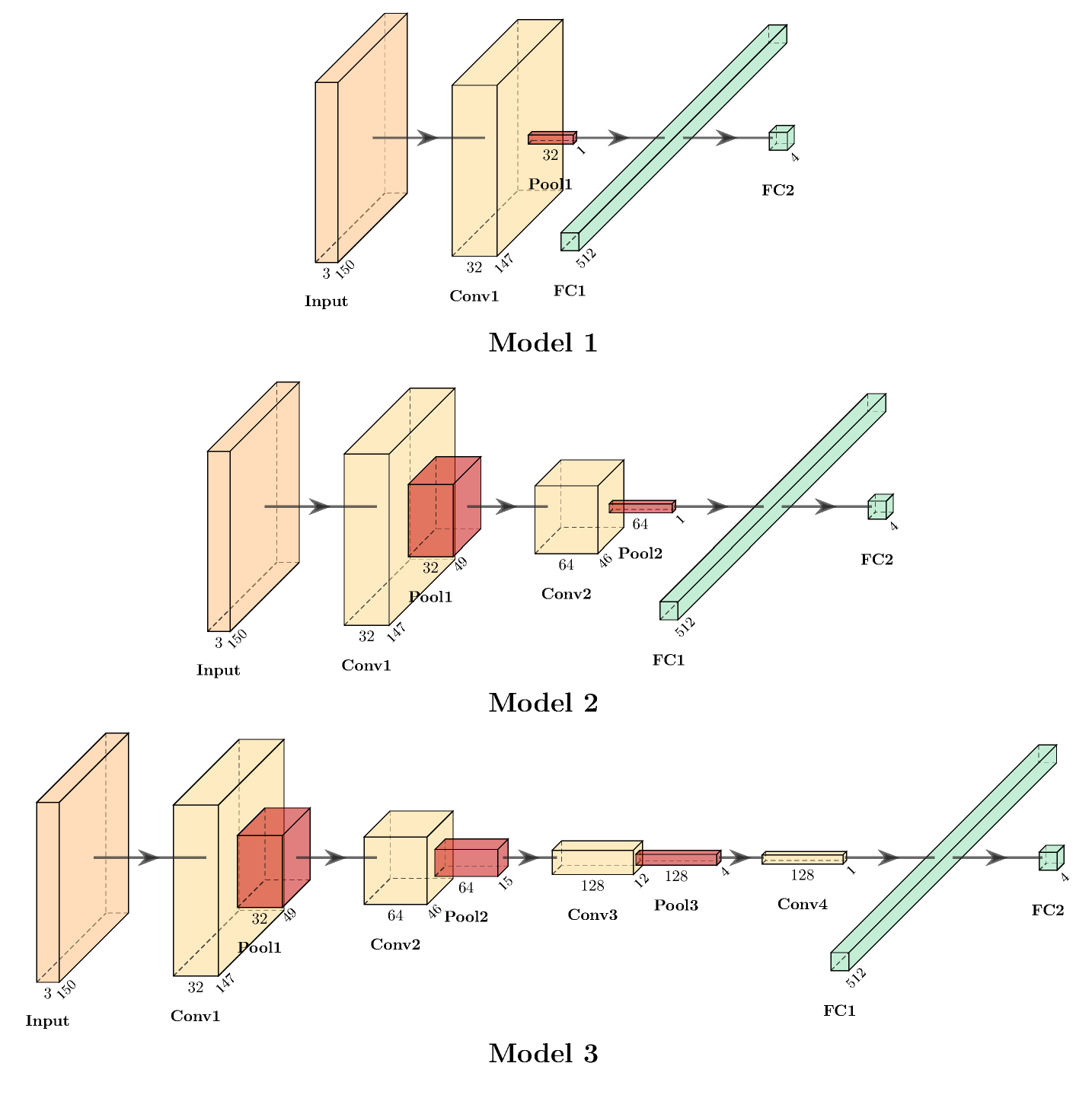}
    \caption{Three CNN architectures with increasing number of layers.}
    \label{fig:structure}
\end{figure}

In the study, we follow the earlier practice and partition the dataset with an approximate 4:1 ratio into a training set with 5,712 samples for model development, and a test set with 1,311 samples for model validation.
To investigate the impact of feature extraction capabilities caused by different depths on model performance in deep learning,
we design three CNN models with an increasing number of convolutional layers, learning features hierarchically from low-level to high-level:
\begin{itemize}
\item[] \textbf{Model 1:} consists of 3 blocks (1 convolutional block and 2 fully-connected blocks). The convolutional block consists of a convolutional layer and a global max-pooling layer.

\item[] \textbf{Model 2:} consists of 4 blocks (2 convolutional blocks and 2 fully-connected blocks). Each convolutional block consists of a convolutional layer and a max-pooling layer (Max-pooling is global in the second convolutional block).

\item[] \textbf{Model 3:} consists of 6 blocks (4 convolutional blocks and 2 fully-connected blocks). All convolutional blocks consist of a convolutional layer and a max-pooling layer except for the fourth, which is only made up of a single convolutional layer.
\end{itemize}
The specific structure of three CNN models with increasing depths is presented in Figure \ref{fig:structure}.
The brain tumor images are adjusted to 150 $\times$ 150 $\times$ 3 dimension as the input to three models, which can be viewed as a stack of 150 by 150 size images with 3 channels. 
Throughout the three deep learning models 
we apply filter size of 4 $\times$ 4 and max-pooling size of 3 $\times$ 3 (global max-pooling not included) with all strides equal to 1.
ReLU is employed as the activation function except in the last fully-connected block, where softmax is used to obtain probability vectors for final classification.
Adam optimizer with learning rate 0.001 is adopted in training.
Models are all developed with the epoch of 40 and batch size of 32 in Python 3.11.

We apply our developed formula to calculate the PDI and its category-specific components for the test sample data in order to assess the discriminative ability of the three CNN models. 
The asymptotic standard errors of the PDI are obtained by applying the fast formula developed in \eqref{hatsigma2notie} and  
95\% asymptotic confidence intervals (CIs) are then constructed. 

Results are summarized in Table \ref{RA}. 
Remarkable improvement in the discriminative ability of the CNN models is observed with the growing depths of the learning structure. The  PDI's increase from 72.9\% (CI: 70.9\%$-$74.9\%) for Model 1 (moderately accurate), followed by 97.3\% (CI: 96.6\%$-$98.0\%) for Model 2 (very good), to 99.7\% (CI: 94.4\%$-$100\%) for Model 3 (almost perfect).
The component-wise accuracy shows that meningioma is the most likely to be misclassified with the lowest accuracy estimates among four categories in all three models of 57.2\% (CI: 52.6\%$-$61.8\%), 92.4\% (CI: 90.3\%$-$94.5\%), and 99.4\% (CI: 98.7\%$-$100\%).
Specifically, Model 3 demonstrates a strong capability in classifying brain tumor MRI images with all component-wise point estimates exceeding 99.0\%. The no-tumor group attains an estimated accuracy of 100\% (CI: 100\%$-$100\%) which means a perfect specificity for these healthy control subjects, followed by glioma of 99.9\% (CI: 99.8\%$-$100\%),
pituitary of 99.6\% (CI: 99.0\%$-$100\%), and meningioma of 99.4\% (CI: 98.7\%$-$100\%). 
The strong performance of deep convolutional neural network (CNN) models in classifying brain tumor MRI images across four diagnostic categories demonstrates their promising potential as tools to assist clinical decision-making. To ensure generalizability and robustness, further validation using external and independent datasets is warranted. To support such efforts, our proposed fast calculation formula for PDI and its variance can be easily implemented. Our development in this paper thus serves as an effective tool for evaluating multi-class diagnostic performance in these complex classification tasks.

\begin{table}[htbp]
\centering
\caption{Estimates and 95\% CI of
the PDI and its components for discriminating brain tumors.}
\label{RA}
\begin{threeparttable}
\begin{tabular}{@{}cccc@{}}
\toprule
 & \textbf{Model 1} & \textbf{Model 2} & \textbf{Model 3} \\
\midrule
Overall PDI & 72.9\% (70.9\%$-$74.9\%) & 97.3\% (96.6\%$-$98.0\%) & 99.7\% (99.4\%$-$100\%) \\
\addlinespace[0.5em]
Category specific accuracy&&&\\
\addlinespace[0.3em]
Glioma & 72.4\% (68.5\%$-$76.3\%) & 98.3\% (97.6\%$-$99.0\%) & 99.9\% (99.8\%$-$100\%) \\
\addlinespace[0.1em]
Meningioma & 57.2\% (52.6\%$-$61.8\%) & 92.4\% (90.3\%$-$94.5\%) & 99.4\% (98.7\%$-$100\%) \\
\addlinespace[0.1em]
No tumor & 91.3\% (89.2\%$-$93.4\%) & 99.0\% (98.4\%$-$99.6\%) & 100\% (100\%$-$100\%) \\
\addlinespace[0.1em]
Pituitary & 70.6\% (66.5\%$-$74.7\%) & 99.5\% (99.2\%$-$99.8\%) & 99.6\% (99.0\%$-$100\%) \\
\bottomrule
\end{tabular}
\begin{tablenotes}
\footnotesize
\item Abbreviations: PDI, polytomous discrimination index.
\end{tablenotes}
\end{threeparttable}
\end{table}

\section{Discussion}
The PDI provides a comprehensive measure of classifier performance derived directly from predicted risk probability vectors. 
This property contributes to PDI's growing adoption in classification problems, 
particularly within machine learning applications. 
Our work establishes a closed-form asymptotic variance estimator for PDI using $U$-statistic theory \cite{Lehmann1999}. 
This solution overcomes computational bottlenecks inherent in bootstrap-dependent methods, 
enabling efficient uncertainty quantification for large-scale multi-class problems without resampling. 
The estimator explicitly accommodates tied observations,
which frequently occur in discrete classification outputs \cite{feng2021discrete}.

There remains substantial scope for the application of multi-class models and PDI evaluation in clinical medicine 
(cf. \cite{Biesheuvel2008b,vancalster2026enemies}). 
A central consideration is that outcome categories should reflect clinically meaningful distinctions, 
particularly when different disease subtypes necessitate different management strategies. 
Representative examples include the ADNEX model \cite{VanCalster2014} for preoperative risk assessment of ovarian cancer; 
ultrasound-based models for abnormal uterine bleeding \cite{Wynants2022}; 
early pregnancy outcome prediction \cite{VanCalster2016a} based on initial serum progesterone and serial serum hCG measurements; 
prediction algorithms aimed at improving early cancer diagnosis \cite{HippisleyCox2025}; 
and machine learning approaches for brain disease classification \cite{Sreedevi2025}, 
such as ensemble convolutional neural networks with Bayesian learning. 
In addition, multi-state prognostic models \cite{Jiang2022} and competing risks frameworks \cite{Ding2021} 
further illustrate the breadth of multi-category modeling approaches for time-dependent accuracy analysis in clinical research.

Quality statistical methods for PDI applications need to be developed, 
including the sample size determination \cite{Pate2023},  
discrimination and calibration assessment methods \cite{Calstervlbts2012,VanHoorde2014}, among others.  
This paper contributes to this evolution by facilitating standard error and confidence interval calculations for the PDI, 
a multi-class $c$-statistic. Adapting the framework to ordinal outcomes using position-weighted PDI variants 
would broaden clinical utility where disease severity gradations exist. 
An ordinal C-index (ORC) was proposed earlier in Van Calster et al. \cite{vanCalster2012discrimination} 
and could serve as a basis for this further development. 

\bibliographystyle{unsrt}
\bibliography{ref}

\end{document}